\documentclass[aps,pra,reprint,amsmath,amssymb,superscriptaddress,nobibnotes, longbibliography]{revtex4-1}
\usepackage{graphicx,SIunits}
\usepackage{bm}     
\usepackage{hyperref} 
\usepackage{dsfont}    
\usepackage{color}

\usepackage{amsmath}
\usepackage{amsthm}
\usepackage{amssymb}
\usepackage{tikz}

\usetikzlibrary{decorations.pathreplacing}
\usepackage{braket}
\usepackage{dsfont}

\usepackage{color}
\usepackage{lineno}
\usepackage{etoolbox}

\usepackage{multirow}

\begin{document}

\preprint{APS/123-QED}

\title{Universal Operational Privacy in Distributed Quantum Sensing}

  \author{Min Namkung}
  \email{These authors have contributed equally.}
    \affiliation{Center for Quantum Technology, Korea Institute of Science and Technology (KIST), Seoul 02792, Korea}

  \author{Dong-Hyun Kim}
  \email{These authors have contributed equally.}
    \affiliation{Center for Quantum Technology, Korea Institute of Science and Technology (KIST), Seoul 02792, Korea}
    \affiliation{Department of Physics, Yonsei University, Seoul 03722, Korea}

    \author{Seongjin Hong}
    \affiliation{Department of Physics, Yonsei University, Seoul 03722, Korea} 

    \author{Yong-Su Kim}
    \affiliation{Center for Quantum Technology, Korea Institute of Science and Technology (KIST), Seoul 02792, Korea}
    \affiliation{Division of Quantum Information, KIST School, Korea University of Science and Technology, Seoul 02792, Korea}

    \author{Su-Yong Lee}
    \affiliation{Emerging Science and Technology Directorate, Agency for Defense Development, Daejeon, 34186, Korea}
    \affiliation{Weapon Systems Engineering, ADD School, University of Science and Technology, Daejeon, 34060, Korea}

  \author{Hyang-Tag Lim}
      \email{forestht@gmail.com}
      \affiliation{Center for Quantum Technology, Korea Institute of Science and Technology (KIST), Seoul 02792, Korea}
      \affiliation{Division of Quantum Information, KIST School, Korea University of Science and Technology, Seoul 02792, Korea}

\date{\today}

\begin{abstract}
Privacy is a fundamental requirement in distributed quantum {{sensor network}}s, where multiple clients estimate spatially distributed parameters using shared quantum resources while interacting with potentially untrusted servers. Despite its importance, existing privacy conditions rely on idealized quantum bounds and do not fully capture the operational constraints imposed by realistic measurements. Here, we introduce a universal operational privacy framework for distributed quantum sensing, formulated in terms of the experimentally accessible Fisher information matrix and applicable to arbitrary protocols characterized by singular information structures. The proposed condition provides a protocol-independent criterion ensuring that no information about individual parameters is accessible to untrusted parties. We further experimentally demonstrate that a distributed quantum sensing protocol employing fewer photons than the number of estimated parameters simultaneously satisfies the universal privacy condition and achieves Heisenberg-limited precision. Our results establish universal operational constraints governing privacy in distributed quantum {{sensor network}}s and provide a foundation for practical, privacy-preserving quantum sensing beyond full-rank regimes.
\end{abstract}


\maketitle

\section{Introduction}
Quantum-enhanced multiple parameter estimation exploits quantum probe states and collective measurements to surpass classical precision limits~\cite{r.demkowicz0,m.szczy,e.polino,m.barbieri,u.doner,r.demkowicz2,l.hwang}. Such quantum advantages have been demonstrated in a variety of sensing schemes, where quantum resources outperform classical strategies in the simultaneous estimation of multiple parameters~\cite{p.c.humphreys,j.-d.yue,e.roccia,j.urrehman,m.namkung,m.namkung2}, including experimental implementations in photonic platforms~\cite{e.polino2,m.kacprowicz,s.hong,s.hong2}. These ideas naturally extend to distributed quantum sensing, in which a global function of spatially distributed parameters is estimated using shared quantum probe states and local measurement devices~\cite{m.gessner,t.j.proctor,w.ge,z.zhang,l.-z.liu,s.-r.zhao,m.malitesta,d.-h.kim,l.pezze,d.-h.kim2}. Distributed quantum sensing has broad relevance across emerging quantum technologies~\cite{f.albarelli,v.montenegro}, including gravitational-wave detection~\cite{r.demkowicz_g}, clock synchronization~\cite{r.k.gosalia}, and remote object sensing~\cite{z.huang}. Despite these advances, the role of privacy as a fundamental constraint in distributed quantum sensing has yet to be fully clarified.

In realistic distributed {{sensor network}}s, privacy becomes a critical requirement: clients may wish to estimate global properties of distributed parameters without revealing individual parameter values to untrusted servers or eavesdroppers~\cite{h.-l.huang,b.polacchi,y.-c.wei}. This has led to the concept of private distributed quantum sensing, where probe states are engineered such that local parameters cannot be independently inferred~\cite{n.shettel,j.dejong,m.hassani,l.bugalho,a.junior,j.ho,u.alushi}. Existing privacy conditions are typically established using probe states associated with singular quantum Fisher information matrices (QFIMs). In particular, Greenberger--Horne--Zeilinger (GHZ) states provide a rank-1 QFIM that guarantees privacy~\cite{m.hassani,l.bugalho,a.junior,j.dejong}, a property that has been theoretically verified using continuity relations of the QFIM~\cite{m.hassani,a.t.rezakhani} and experimentally demonstrated~\cite{j.ho}. These approaches implicitly equate privacy with rank deficiency of the {QFIM}, assuming that ideal quantum bounds faithfully represent the information accessible to untrusted servers.

Despite these advances, existing continuity-based privacy conditions remain restricted to rank-1 QFIMs and do not generalize to realistic distributed quantum sensing scenarios. In practice, distributed sensing schemes often involve higher-rank singular Fisher information matrices (FIMs), especially in multiparameter estimation settings or under resource constraints~\cite{a.junior}. Moreover, attaining the QFIM itself can be experimentally challenging, as optimal measurements may require infeasible entangling operations~\cite{p.c.humphreys,s.hong,s.hong2} or arise from parameters encoded via incompatible Hamiltonians~\cite{y.yang}. This reveals a fundamental mismatch between existing privacy conditions, formulated at the level of the QFIM, and the operational notion of privacy determined by the information actually extracted by realistic measurements. These limitations call for a universal privacy principle formulated directly in terms of experimentally accessible quantities. {Moreover, when concerning FIM having rank more than one, it is particularly necessary that the privacy condition clarifies whether even a single parameter is protected against any untrusted parties~\cite{a.junior}, eventually further improving the privacy.}

In this work, we identify a universal operational framework for privacy in distributed quantum sensing, formulated in terms of the experimentally accessible FIM. The proposed framework provides a protocol-independent criterion that simultaneously characterizes the information accessible to untrusted servers and the precision attainable by clients, and applies to arbitrary distributed quantum sensing protocols characterized by singular information structures with rank greater than one. We further experimentally demonstrate that a distributed quantum sensing protocol employing fewer photons than the number of estimated parameters~\cite{d.-h.kim} satisfies the universal privacy condition while achieving Heisenberg-limited precision~\cite{j.wang,m.namkung3}. Our results uncover universal operational constraints governing privacy in distributed quantum {{sensor network}}s and provide a foundation for practical, privacy-preserving quantum sensing beyond full-rank sensing regimes.

\begin{figure}
\centerline{\includegraphics[width=\columnwidth]{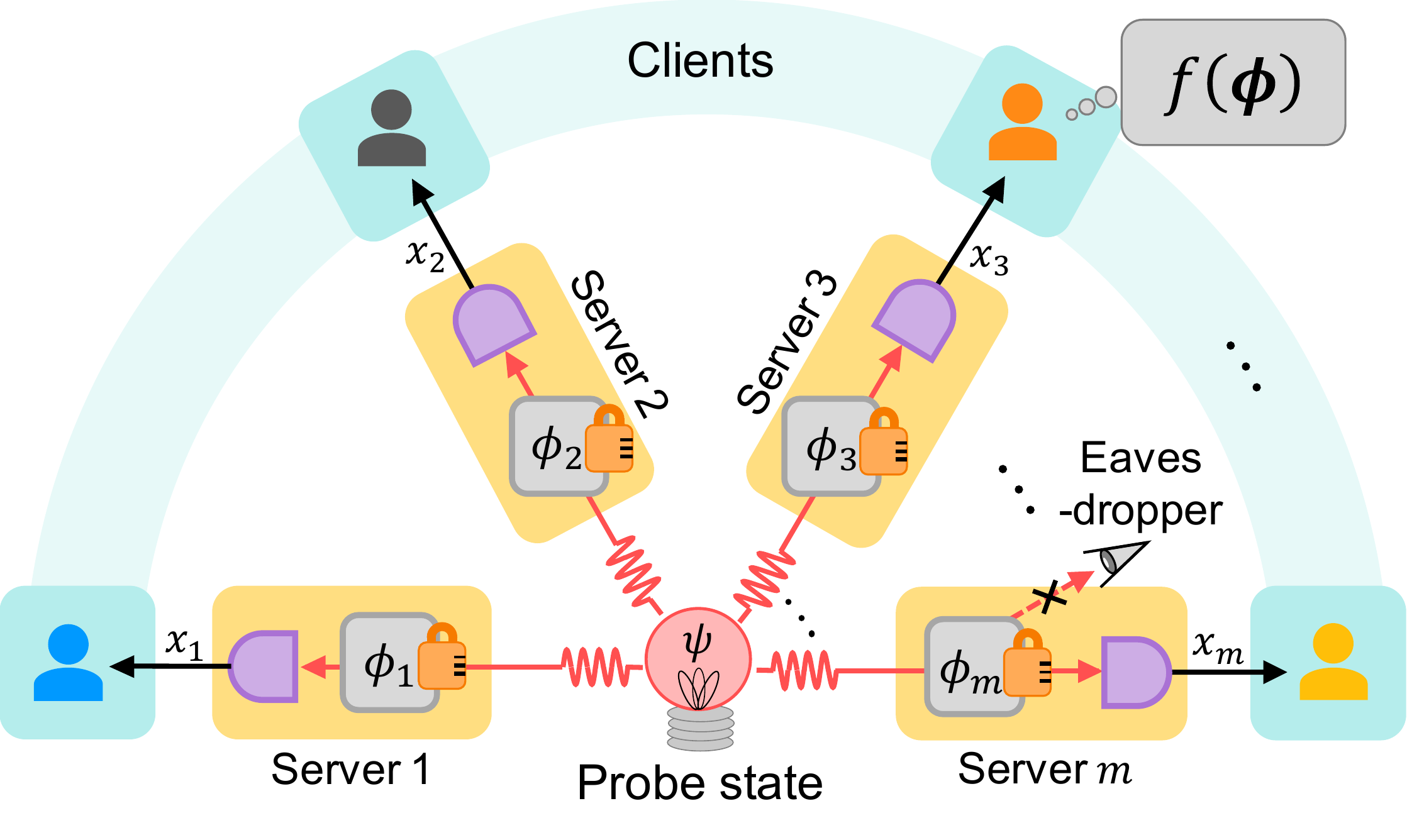}}
\caption{Conceptual schematic of a private distributed quantum {{sensor network}}. A shared quantum probe state is encoded with spatially distributed parameters $\bm{\phi}$ such that no information about individual parameters is accessible to untrusted servers. The servers perform local measurements on the encoded state, and the resulting outcomes $\{x_1,\ldots,x_m\}$ are communicated to the clients, who estimate a global function $f(\bm{\phi})=\bm{w}^{\mathrm{T}}\bm{\phi}$.}

\centering
\label{fig:1}
\end{figure}

\section{Background}
We consider a general scenario of private distributed quantum sensing, schematically illustrated in Fig.~\ref{fig:1}. A probe state described by a density operator $\hat{\rho}$, which may be taken as a pure state $|\psi\rangle\langle\psi|$, is encoded with a set of parameters $\bm{\phi}=\{\phi_1,\cdots,\phi_m\}$ in a parameter space $\mathbb{R}^{m}$ through a unitary operation $\hat{U}_{\bm{\phi}}$. The parameter-encoded state $\hat{\rho}_{\bm{\phi}}=\hat{U}_{\bm{\phi}}\hat{\rho}\hat{U}_{\bm{\phi}}^{\dagger}$ is distributed to spatially separated servers, which perform local measurements using feasible measurement devices. Based on the measurement outcomes $\bm{x}=\{x_1,\cdots,x_m\}$, clients estimate a global linear function of the parameters, $f(\bm{\phi})=\bm{w}^{\mathrm{T}}\bm{\phi}$~\cite{l.pezze_ph}, while ensuring that the individual local parameters are not revealed to untrusted servers.

Within the quantum metrological framework~\cite{c.w.helstrom}, the estimation precision is characterized by the {FIM} $\mathbf{F}$, whose elements are defined as 

\begin{equation*}
\mathbf{F}_{\mu\nu}={\sum_{\bm{x}}\frac{1}{p(\bm{x}|\bm{\phi})}\frac{\partial  p(\bm{x|\bm{\phi}})}{\partial{\phi_\mu}}\frac{\partial  p(\bm{x|\bm{\phi}})}{\partial\phi_\nu}},
\end{equation*}
where $p(\bm{x}|\bm{\phi})=\mathrm{Tr}(\hat{\rho}_{\bm{\phi}}\hat{M}_{\bm{x}})$ is the probability of obtaining outcome $x$ from a positive operator-valued measure (POVM) $\{ \hat{M}_{\bm{x}}\}$ and $\mu,\nu\in\{1,2,\cdots,m\}$~\cite{h.cramer,c.r.rao}. The {FIM} is upper bounded by the {QFIM} $\mathbf{Q}$, whose elements are given by
\begin{equation*}
\mathbf{Q}_{\mu\nu}=\frac{1}{2}\mathrm{Tr}\big(\hat{\rho}\big\{\hat{L}_\mu,\hat{L}_\nu\big\}\big),
\end{equation*}
with $\hat{L}_\mu$ denoting the symmetric logarithmic derivative with respect to $\phi_\mu$~\cite{c.w.helstrom,j.liu,m.g.a.paris}. As a result, the inequality
\begin{equation*}
\bm{w}^{\rm T}\mathbf{F}\bm{w}\le\bm{w}^{\rm T}\mathbf{Q}\bm{w}
\end{equation*}
holds for any weight vector $\bm{w}$.

{We define privacy operationally as the impossibility of inferring any individual parameter from the full classical data accessible to untrusted servers, quantified through the structure of the rank-deficient FIM~\cite{m.hassani,a.junior}. This can be justified as follows: if the FIM is supposed to be invertible, then it allows estimating arbitrary weighted function $\bm{w}^{\rm T}\bm{\phi}$, which includes individually estimating local parameters $\phi_1,\cdots,\phi_m$~\cite{footnote}.}  In this regime, the precision of estimating a global function $\bm{w}^{\rm T}\bm{\phi}${, quantified by a variance $\Delta^2(\bm{w}^{\rm T}\bm{\phi})$,} is bounded by the (quantum) Cram\'{e}r-Rao inequality
\begin{equation*}
    \Delta^2(\bm{w}^{\rm T}\bm{\phi})\ge\bm{w}^{\rm T}\mathbf{F}^+\bm{w}\ge\bm{w}^{\rm T}\mathbf{Q}^+\bm{w},
\end{equation*}
where the superscript ``$+$'' denoting the Moore-Penrose pseudoinverse~\cite{r.penrose}. These bounds reflect the fact that privacy arises precisely when certain parameter directions remain inaccessible to the servers' measurements.

More generally, when estimating multiple global functions $\bm{w}^{(1)\mathrm{T}}\bm{\phi},\bm{w}^{(2)\mathrm{T}}\bm{\phi},\cdots,\bm{w}^{(r)\mathrm{T}}\bm{\phi}$, or when estimating all of them simultaneously, the attainable lower bound on the total variance is expressed as
\begin{equation*}
    \sum_{j}\Delta^2(\bm{w}^{(j)\mathrm{T}}\bm{\phi})\ge\mathrm{Tr}\left\{\mathbf{W}\mathbf{F}^+\right\}\ge\mathrm{Tr}\left\{\mathbf{W}\mathbf{Q}^+\right\},
\end{equation*}
where $\mathbf{W}=\sum_{j}\bm{w}^{(j)}\bm{w}^{(j)\mathrm{T}}$ is the corresponding weight matrix~\cite{t.j.proctor,a.fujiwara,m.g.genoni,c.vaneph}. These bounds are saturable when the weight vectors lie within the support space of the {QFIM}, defined by the eigenvectors associated with nonzero eigenvalues of $\mathbf{Q}$~\cite{m.namkung3}. \\

\begin{figure*}[t]
\includegraphics[width=\textwidth]{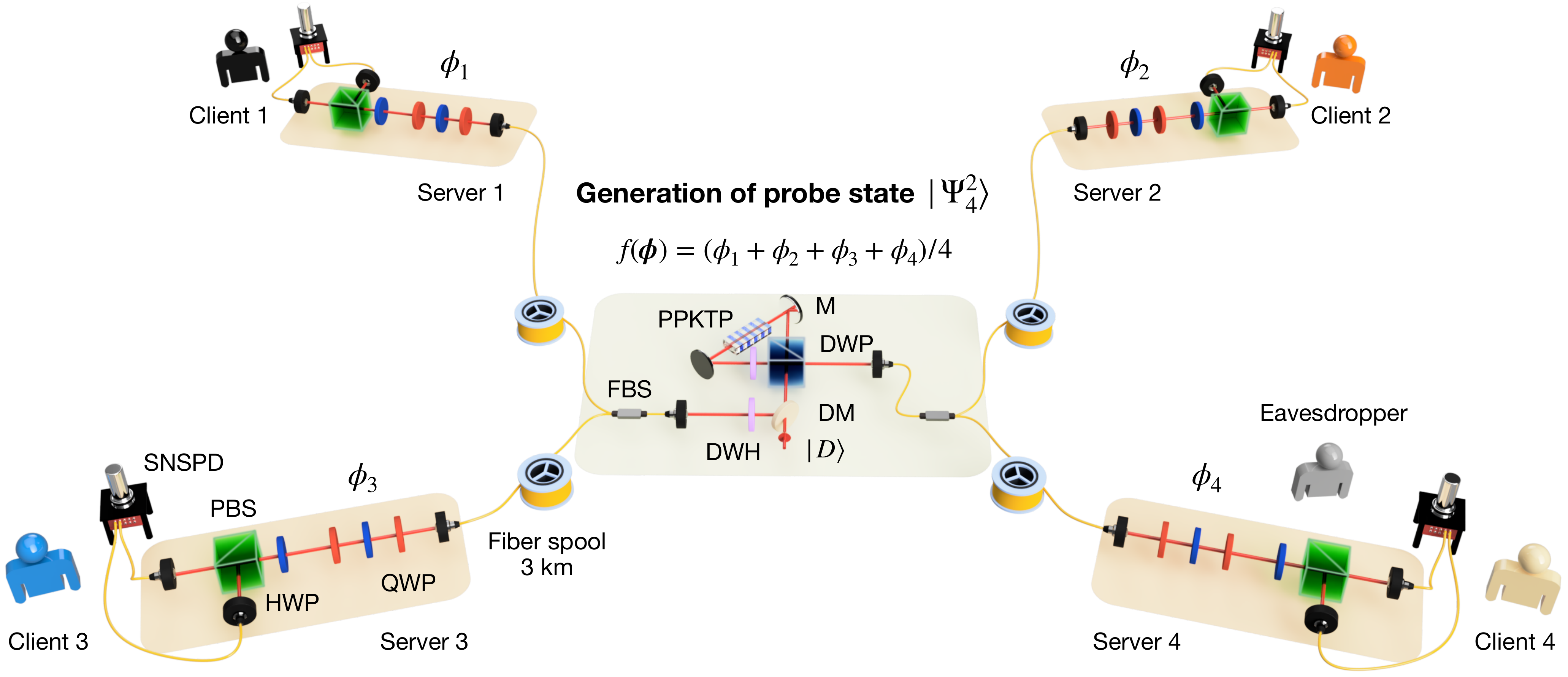} 
\caption{Experimental implementation of a private distributed quantum {{sensor network}}. A polarization-entangled Bell state is generated in a Sagnac interferometer and distributed through a beam splitter network (BSN) to prepare the two-photon state $|\Psi_4^2\rangle$. The state is sent to spatially separated servers via 3-km optical fiber links, where local phase encoding $\hat{U}(\bm{\phi})$ is applied. Each server performs projective measurements in the $\hat{\sigma}_x$, and two-photon coincidence events are recorded and transmitted to the clients for estimating global functions of the distributed phases.  SNSPD: superconducting nanowire single-photon detector; QWP: quarter waveplate; HWP: half waveplate; PBS: polarizing beam splitter; DM: dichroic mirror; DWP: dual wavelength PBS; DWM: dual wavelength mirror; DWH: dual wavelength HWP; PPKTP: periodically-poled KTiOPO$_4$; BSN: beam splitter network; FBS: $50/50$ fiber beam splitter.}
\label{fig:figure2}
\end{figure*}

\section{Results}

\subsection{Universal privacy condition}
A central challenge in private distributed quantum sensing is to certify privacy under realistic measurement conditions, where the QFIM may not be experimentally attainable~\cite{p.c.humphreys,s.hong,s.hong2,y.yang}. Since privacy is ultimately determined by the information actually extracted by measurements performed by untrusted servers, a meaningful privacy condition must be formulated directly in terms of the experimentally accessible FIM and remain valid for arbitrary preparation and measurement configurations.

Motivated by this operational requirement, we introduce a universal privacy quantifier defined for a weight vector $\bm{w}$ as
\begin{equation}
\mathcal{P}_{\mathbf{F}}(\bm{w})
=
1-\min_{\bm{v}\perp\bm{w},\,\|\bm{v}\|_2=1}
\bm{v}^{\mathrm{T}}\mathbf{\Pi}_{\mathbf{F}}\bm{v},
\label{pp}
\end{equation}
where $\mathbf{\Pi}_{\mathbf{F}}$ denotes the projector onto the support space of the FIM $\mathbf{F}$, spanned by the eigenvectors associated with nonzero eigenvalues. {Here, the vector $\bm{v}$ is constrained to orthogonal to $\bm{w}$, such that the value of $\mathcal{P}_{\mathbf{F}}(\bm{w})$ becomes zero when $\bm{\Pi}_{\mathbf{F}}$ has a kernel space, meaning that $\mathbf{F}$ is rank-deficient.} {This covers the case of GHZ states when $\Pi_{\mathbf{F}}$ is a rank-1 projector~\cite{m.hassani}.} Moreover, the quantifier satisfies $0\le\mathcal{P}_{\mathbf{F}}(\bm{w})\le1$ and can be generalized to the simultaneous estimation of multiple global functions. {The explicit generalization of the privacy quantifier to multiple functions is provided as
\begin{equation}
{\mathcal{P}_{\mathbf{F}}(\mathbf{W})
=
1-\frac{1}{r}
\min_{\substack{\bm{v}^{(j)}\perp\bm{w}^{(j)}\\ \|\bm{v}^{(j)}\|_2=1}}
\sum_{j=1}^{r}
\bm{v}^{(j)\mathrm{T}}\mathbf{\Pi}_{\mathbf{F}}\bm{v}^{(j)},}
\label{gen_p}
\end{equation}
{where $r$ denotes the rank of the {FIM} and $\bm{w}^{(j)}$ are weight vectors constituting the weight matrix $\mathbf{W}$. The detailed derivation is contained} in {Appendix A}. {The above extension suggests that it is possible to privately estimate multiple global functions simultaneously, improving versatility of the private {sensor network}.}}

Geometrically, $\mathcal{P}_{\mathbf{F}}(\bm{w})$ characterizes how strongly directions in parameter space orthogonal to $\bm{w}$ remain hidden from the servers' measurements. If the FIM is non-singular, $\mathbf{\Pi}_{\mathbf{F}}$ acts as the identity on the entire parameter space, yielding $\mathcal{P}_{\mathbf{F}}(\bm{w})=0$ and indicates the absence of privacy. By contrast, when the FIM is singular, the support space becomes a strict subspace of the parameter space, leading to $\mathcal{P}_{\mathbf{F}}(\bm{w})>0$ and the emergence of privacy. In the extreme case where the kernel of the FIM contains directions fully orthogonal to $\bm{w}$, the minimum in Eq.~(\ref{pp}) vanishes and $\mathcal{P}_{\mathbf{F}}(\bm{w})=1$, meaning that both desired privacy and sensitivity are simultaneously achieved by the clients.

\begin{figure*}[t]
\includegraphics[width=\textwidth]{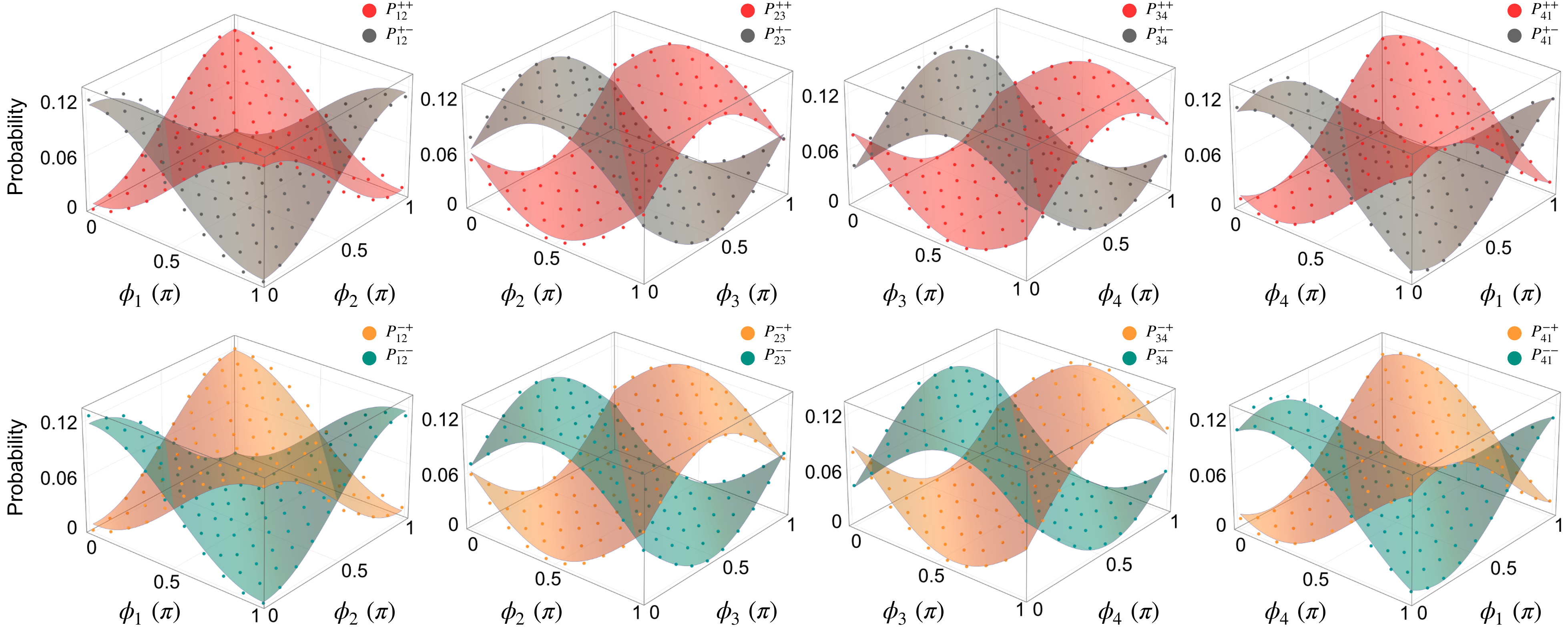} 
\caption{Experimentally measured two-photon outcome probabilities for the distributed quantum sensing protocol. The sixteen probability surfaces are fitted to the experimental data points using the probability functions $P_{\mu\nu}^{\pm\pm}$ and $P_{\mu\nu}^{\pm\mp}$ in Eq.~(\ref{eq:P}), with $\{\mu \nu\}=\{12,23,34,41\}$, obtained by scanning $\phi_1$, $\phi_2$, $\phi_3$, and $\phi_4$ within the interval $[0,\pi]$. Error bars are smaller than the marker size.}
\label{fig:figure3}
\end{figure*}

Importantly, the proposed privacy condition clarifies that not only the exposure of all local parameters but also the partial exposure of any individual parameter is prevented{, which is concerned by the previous work~\cite{a.junior}}. This property arises from the geometric structure of the FIM kernel and is independent of the particular estimation strategy. {For illustration, we consider a {sensor network} tasked with estimating the global function $f(\bm{\phi})=\phi_1+\phi_2+\phi_3+\phi_4$ as an example. The {sensor network} is assumed to be characterized by rank-3 {FIM}, with a kernel space spanned by $\bm{v}_{\rm ker}=[1 \ -1 \ 1 \ -1]$. An adversary aiming to partially infer $\phi_1$ and $\phi_2$ may attempt to estimate two additional functions $g(\bm{\phi})=\phi_1+a\phi_2+b(\phi_3+\phi_4)$ and $h(\bm{\phi})=\phi_1+c\phi_2+d(\phi_3+\phi_4)$, without loss of generality. To make them possible, the associated weight vectors $[1 \ \ a \ \ b \ \ b]$ and $[1 \ \ c \ \ d \ \ d]$ must lie in the space perpendicular to $\bm{v}_{\rm ker}$. This condition eventually admits infinitely many solutions of the following equations:
\begin{eqnarray}
    \begin{bmatrix}
        f(\bm{\phi})\\
        g(\bm{\phi})\\
        h(\bm{\phi})
    \end{bmatrix}=\begin{bmatrix}
        1& 1& 1\\
        1& a& b\\
        1& c& d
    \end{bmatrix}
    \begin{bmatrix}
        \phi_1\\
        \phi_2\\
        \phi_3+\phi_4
    \end{bmatrix},
\end{eqnarray}
thereby preventing an untrusted party to estimate $\phi_1$ and $\phi_2$. This observation can be naturally incorporated into our framework. When $\bm{v}_{\rm ker}$ determines the privacy measure $\mathcal{P}(\bm{w})$ of Eq.~(\ref{pp}), it ensures that no individual local parameters is accessible to any untrusted party, even when the {FIM} has rank greater than one.} A detailed discussion illustrating how the kernel structure prevents the exposure of any local parameter is presented in {Appendix B}.

The operational definition of privacy revealed by $\mathcal{P}_{\mathbf{F}}(\bm{w})$ also elucidates the interplay between privacy and attainable precision. When the direction minimizing Eq.~(\ref{pp}) has finite overlap with the support space of the FIM, $\mathcal{P}_{\mathbf{F}}(\bm{w})<1$, indicating that privacy is preserved at the expense of reduced sensitivity. This regime reflects a geometric trade-off arising from the structure of the FIM, whereby a probe state can protect all local parameters while preventing the global function from saturating the Cram\'er--Rao bound~\cite{m.namkung3,y.-h.li}. The geometric interpretation of the intermediate regime $0<\mathcal{P}_{\mathbf{F}}(\bm{w})<1$ is discussed in {Appendix C}.

The proposed privacy quantifier $\mathcal{P}_{\mathbf{F}}(\bm{w})$ satisfies several properties essential for physical validity and experimental feasibility {as discussed in} {Appendix D}. It is invariant under changes of basis in the parameter space, obeys a continuity relation under small experimental imperfections, and is robust against parameter-independent noise~\cite{a.junior}. In particular, for an experimentally reconstructed FIM $\mathbf{F}(\epsilon)$ deviating slightly from the ideal $\mathbf{F}$ by an amount $\epsilon\ll1$, the privacy quantifier satisfies
\begin{equation}
\big|\mathcal{P}_{\mathbf{F}(\epsilon)}(\bm{w})-\mathcal{P}_{\mathbf{F}}(\bm{w})\big|
=O(\epsilon^2).
\label{continuity}
\end{equation}
The basis independence, continuity, and noise tolerance of the proposed privacy quantifier are formally verified in Methods. {We note that all these three quantities are compatibly satisfied by the generalized indicator of Eq.~(\ref{gen_p}).} These properties establish $\mathcal{P}_{\mathbf{F}}(\bm{w})$ as a universal operational principle for privacy in distributed quantum sensing.\\

\subsection{Experimental demonstration} 
To experimentally validate the proposed universal operational privacy condition, we implement a distributed quantum {{sensor network}} based on entangled photon pairs, as schematically illustrated in Fig.~\ref{fig:figure2}. The experiment realizes a Heisenberg-scaled sensing protocol employing fewer photons than the number of encoded parameters, while directly demonstrating privacy at the level of the experimentally accessible FIM.

\begin{figure*}[t]
\includegraphics[width=2\columnwidth]{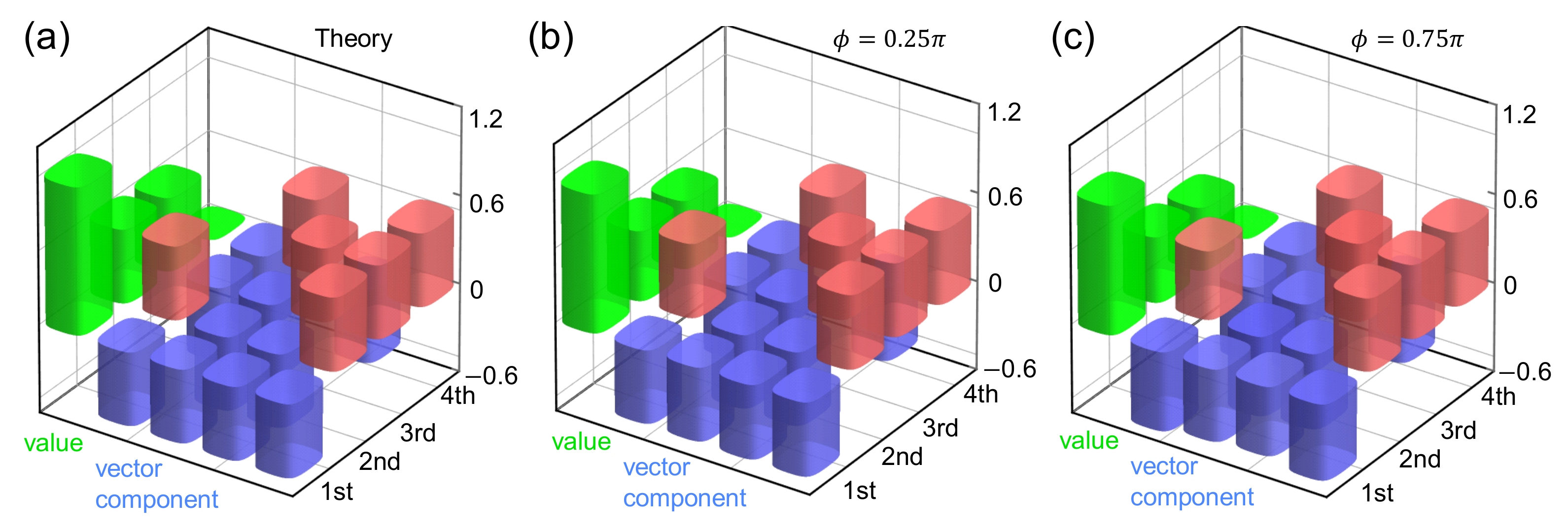} 
\caption{Experimental verification of privacy in distributed quantum sensing via the singular FIM. Eigenvalues and eigenvectors of the FIM, where the horizontal and vertical axes denote eigenvalues and the corresponding vector components, respectively. (a) Theoretical FIM. (b) Experimentally reconstructed FIM at $\phi = 0.25\pi$. (c) Experimentally reconstructed FIM at $\phi = 0.75\pi$.}
\label{fig:figure4}
\end{figure*}

A continuous-wave laser at a central wavelength of 780.25~nm is prepared in the diagonal polarization state $|D\rangle=(|H\rangle+|V\rangle)/\sqrt{2}$ and injected into a Sagnac interferometer. Polarization-entangled photon pairs are generated via spontaneous parametric down-conversion in a 10-mm-thick type-II periodically poled KTiOPO$_4$ (PPKTP) crystal, yielding the Bell state $|\Psi^+\rangle=(|HV\rangle+|VH\rangle)/\sqrt{2}$. After a dual wavelength half waveplate (DWH) set at $45^\circ$, the state is transformed into $|\Phi^+\rangle=(|HH\rangle+|VV\rangle)/\sqrt{2}$. A beam splitter network (BSN) composed of two 50/50 fiber beam splitters (FBSs) then distributes the photons across four spatial modes, preparing the two-photon entangled state~\cite{d.-h.kim}
\begin{eqnarray}
 	|\Psi_4^2\rangle=\frac{1}{2}\left(
    |\Phi_{1,2}^{2}\rangle+|\Phi_{2,3}^{2}\rangle+|\Phi_{3,4}^{2}\rangle+|\Phi_{4,1}^{2}\rangle
    \right),
 \label{eq:psi}
\end{eqnarray}
where $|\Phi_{\mu,\nu}^{2}\rangle=(|H_\mu H_\nu\rangle+|V_\mu V_\nu\rangle)/\sqrt{2}$ denotes a Bell state shared between the $\mu$th and $\nu$th servers. The state $|\Psi_4^2\rangle$ is distributed to the servers through 3-km optical fiber spools.

Each server locally applies a phase-encoding operation $\hat{U}_{\bm{\phi}}$, implemented using two quarter waveplates (QWPs) set at $45^\circ$ and a half waveplate (HWP), with $\bm{\phi}=\{\phi_1,\phi_2,\phi_3,\phi_4\}$. The encoded state is measured via local projective measurements in the $\hat{\sigma}_x$ basis, realized using a HWP set at $22.5^\circ$ followed by a polarization beam splitter and superconducting nanowire single-photon detectors. Two-photon coincidence events are recorded and communicated to the clients.

From the measurement outcomes, the two-photon coincidence probabilities take the form
\begin{eqnarray}
 	p[(x_\mu,x_\nu)=(\pm,\pm)|\bm{\phi}] &=&
    \frac{1}{16}\!\left[1+V^{\pm\pm}_{\mu\nu}\cos(\phi_\mu+\phi_\nu)\right], \nonumber\\
 	p[(x_\mu,x_\nu)=(\pm,\mp)|\bm{\phi}] &=&
    \frac{1}{16}\!\left[1-V^{\pm\mp}_{\mu\nu}\cos(\phi_\mu+\phi_\nu)\right],\nonumber\\
 \label{eq:P}
\end{eqnarray}
where $V^{\pm\pm}_{\mu\nu}$ ($V^{\pm\mp}_{\mu\nu}$) denotes the two-photon interference visibility for correlated (anticorrelated) outcomes, and
$\{\mu\nu\}=\{12,23,34,41\}$ labels the pairs of spatial modes. Notably, each probability depends only on the sum $\phi_\mu+\phi_\nu$, implying that no individual local parameter can be inferred from any single-server measurement outcome.

Assuming unit visibility, the resulting FIM is
\begin{align}\label{eq:FIM}
 	\mathbf{F}=
    \begin{bmatrix}
 		1/2 & 1/4 & 0   & 1/4 \\
 		1/4 & 1/2 & 1/4 & 0   \\
 		0   & 1/4 & 1/2 & 1/4 \\
 		1/4 & 0   & 1/4 & 1/2
 	\end{bmatrix},
\end{align}
which is singular and yields a Heisenberg-scaled uncertainty of $1/N^2=1/4$ for the global weight vector
$\bm{w}=(1,1,1,1)/4$. For this weight vector, the universal privacy quantifier satisfies
$\mathcal{P}_{\mathbf{F}}(\bm{w})=1$, demonstrating that the clients simultaneously achieve optimal precision and perfect privacy. The kernel of the FIM is spanned by the vector $[1\ -1\ 1\ -1]$, allowing multiple global functions to be estimated without exposing individual local parameters as discussed in {Appendix B}.

Experimentally, we estimate the average phase $(\phi_1+\phi_2+\phi_3+\phi_4)/4$ by scanning each phase over the interval $[0,\pi]$. From the measured sixteen outcome probabilities, we obtain an average visibility of $V=0.968\pm0.003$, as shown in Fig.~\ref{fig:figure3}. These data are used to reconstruct the experimental FIM, which closely reproduces the ideal matrix in Eq.~(\ref{eq:FIM}).

The privacy of the distributed {{sensor network}} is further verified by comparing the eigenvalues and eigenvectors of the experimentally reconstructed FIM with those of the theoretical FIM, as shown in Fig.~\ref{fig:figure4}. We find excellent agreement across the full parameter range, confirming that the experimentally realized {{sensor network}} satisfies the universal privacy condition. Additional experimental results demonstrating the persistence of the singular FIM over the entire parameter space are provided in Fig.~\ref{Sfig:1privacy} of {Appendix E}.

\section{Conclusion}

We have established a \emph{universal privacy condition} for distributed quantum sensing that provides an operational criterion for determining whether individual parameters are protected against untrusted parties. The proposed condition applies to arbitrary singular FIMs, including those arising in realistic experimental implementations where optimal measurements may not be accessible. We experimentally verified this condition using a distributed quantum sensing protocol that employs fewer photons than the number of encoded phases while achieving Heisenberg-scaled precision. These results demonstrate that privacy and quantum-enhanced precision can be simultaneously realized in practical quantum {{sensor network}}s.

Beyond the specific implementation reported here, our framework generalizes previous privacy conditions based on QFIMs~\cite{n.shettel,j.dejong,m.hassani,l.bugalho,a.junior,j.ho,u.alushi} by providing a measurement-independent criterion applicable to general experimental settings. The approach is broadly relevant across quantum sensing platforms, including photonic systems~\cite{e.roccia,e.polino2,m.kacprowicz,d.-h.kim,a.junior}, trapped ions~\cite{j.bate}, and superconducting circuits~\cite{j.zhang}. By unifying distributed sensing, multiparameter estimation with rank-greater-than-one singular FIMs, and experimentally accessible privacy verification, our work establishes a general foundation for scalable and privacy-preserving quantum {{sensor network}}s.

\section*{Data availability}
The data that support the findings of this study are available from the corresponding authors upon reasonable request.

\begin{widetext}

\section*{Appendix A. Generalized universal privacy condition}

{Here we extend the universal operational privacy condition introduced in the main text to the case where clients estimate multiple global functions 
$\bm{w}^{(1)\mathrm{T}}\bm{\phi},\ldots,\bm{w}^{(r)\mathrm{T}}\bm{\phi}$ simultaneously.
The corresponding weight vectors $\bm{w}^{(1)},\ldots,\bm{w}^{(r)}\in\mathbb{R}^m$ define a weight matrix
$\mathbf{W}=\sum_{j=1}^{r}\bm{w}^{(j)}\bm{w}^{(j)\mathrm{T}}$.
The generalized privacy quantifier is defined as
\begin{equation}
\mathcal{P}_{\mathbf{F}}(\mathbf{W})
=
1-\frac{1}{r}
\min_{\substack{\bm{v}^{(j)}\perp\bm{w}^{(j)}\\ \|\bm{v}^{(j)}\|_2=1}}
\sum_{j=1}^{r}
\bm{v}^{(j)\mathrm{T}}\mathbf{\Pi}_{\mathbf{F}}\bm{v}^{(j)}.
\label{gen_p_}
\end{equation}}

{This definition consistently extends the operational interpretation of privacy discussed in the main text.
If the FIM $\mathbf{F}$ is non-singular, the projector
$\mathbf{\Pi}_{\mathbf{F}}$ acts as the identity on the entire parameter space.
In this case, each term
$\bm{v}^{(j)\mathrm{T}}\mathbf{\Pi}_{\mathbf{F}}\bm{v}^{(j)}=1$,
leading to $\mathcal{P}_{\mathbf{F}}(\mathbf{W})=0$ and indicating the absence of privacy.}

{By contrast, when $\mathbf{F}$ is singular, $\mathbf{\Pi}_{\mathbf{F}}$ projects onto a strict subspace of the parameter space.
Consequently, $\bm{v}^{(j)\mathrm{T}}\mathbf{\Pi}_{\mathbf{F}}\bm{v}^{(j)}<1$ for all admissible vectors, yielding
$\mathcal{P}_{\mathbf{F}}(\mathbf{W})>0$.
Perfect privacy, $\mathcal{P}_{\mathbf{F}}(\mathbf{W})=1$, is achieved when all minimizing vectors $\bm{v}^{(j)}$ lie entirely within the kernel of $\mathbf{F}$.}

\section*{Appendix B. Principle of no local parameter exposure}

{This subsection clarifies how the proposed privacy condition prevents the exposure of any individual local parameter, even when multiple global functions are estimated.
We consider the estimation of the balanced global function
$f(\bm{\phi})=\bm{w}^{\mathrm{T}}\bm{\phi}=\phi_1+\phi_2+\cdots+\phi_m$ with respect to $\bm{w}_j=1$,
as discussed in previous works~\cite{m.hassani,l.bugalho,a.junior}.
Suppose that the kernel of the FIM contains the vector
$\bm{v}_{\mathrm{ker}}=[1,-1,1,-1,\ldots,1,-1]$.}

{For illustration, let $m=8$ and assume that the following four global functions can be estimated with vanishing variance:
\begin{align}\label{ex}
    {g_2(\bm{\phi})}&=\phi_1+a\phi_2+b\phi_3+c\phi_4+d(\phi_5+\phi_6+\phi_7+\phi_8),\nonumber\\
    {g_3(\bm{\phi})}&=\phi_1+e\phi_2+f\phi_3+g\phi_4+h(\phi_5+\phi_6+\phi_7+\phi_8),\nonumber\\
    {g_4(\bm{\phi})}&=\phi_1+k\phi_2+l\phi_3+m\phi_4+n(\phi_5+\phi_6+\phi_7+\phi_8),\nonumber\\
    {g_5(\bm{\phi})}&=\phi_1+p\phi_2+q\phi_3+r\phi_4+s(\phi_5+\phi_6+\phi_7+\phi_8),
\end{align}
with nonzero coefficients.}

{If the {below equation
\begin{equation}\label{M}
    \begin{bmatrix}
        f(\bm{\phi}) \\
       g_2(\bm{\phi}) \\
       g_3(\bm{\phi}) \\
       g_4(\bm{\phi}) \\
       g_5(\bm{\phi})
    \end{bmatrix}=
    \begin{bmatrix}
    1& 1& 1& 1& 1 \\
        1& a& b& c& d\\
        1& e& f& g& h\\
        1& k& l& m& n\\
        1& p& q& r& s
    \end{bmatrix}
    \begin{bmatrix}
        \phi_1 \\
       \phi_2 \\
       \phi_3 \\
       \phi_4 \\
       \phi_5+\phi_6+\phi_7+\phi_8
    \end{bmatrix}
\end{equation}
has a unique solution}, the local parameters $\phi_1,\phi_2,\phi_3,$ and $\phi_4$ would be uniquely determined and thus exposed.
However, orthogonality to the kernel vector $\bm{v}_{\mathrm{ker}}$ requires all weight vectors in Eq.~(\ref{ex}) to satisfy
$\bm{w}^{(j)}\perp\bm{v}_{\mathrm{ker}}$~\cite{m.namkung,p.stoica,y.-h.li},
which renders the columns of Eq.~(\ref{M}) linearly dependent.
As a result, no individual local parameter can be inferred, even though multiple global functions are estimated.
This argument extends straightforwardly to an arbitrary number of parameters.}

{We further generalize the aforementioned technique to the case of $f(\bm{\phi})=a_{11}\phi_1+\cdots+a_{1m}\phi_m$, together with the consideration of $g_j(\bm{\phi})=a_{j1}\phi_1+\cdots+a_{j  r-1}\phi_{r-1}+a_{1r}\phi_r+\cdots+a_{rm}\phi_m$ with $j=2,\cdots,r$. Let the kernel vector $\bm{v}_{\mathrm{ker}}$ be $\bm{v}_{\mathrm{ker}}=[k_1 \ \ k_2 \ \ \cdots \ \ k_{r-1} \ \ k_r \ \ \cdots \ \ k_r]$ with $k_j\not=0$ for any $j$. To make all $g_j(\bm{\phi})$ be estimated within finite variance, their weight vectors should be orthogonal to $\bm{v}_{\rm ker}$, leading to
\begin{equation}
    k_1\begin{bmatrix} a_{11}  \\ \vdots \\ a_{r1} \end{bmatrix}+k_2\begin{bmatrix} a_{12}  \\ \vdots \\ a_{r2} \end{bmatrix}+\cdots+k_{r-1}\begin{bmatrix} a_{1r-1}  \\ \vdots \\ a_{rr-1} \end{bmatrix}+k_r\begin{bmatrix} 1  \\ \vdots \\ 1 \end{bmatrix}=0.
\end{equation}
This eventually implies that there is no unique solution of the equation
\begin{equation}
    \begin{bmatrix}
        f(\bm{\phi})\\
        g_2(\bm{\phi})\\
        \vdots\\
        g_r(\bm{\phi})
    \end{bmatrix}=\begin{bmatrix}
        a_{11} & a_{12} & \cdots & 1 \\
        a_{21} & a_{22} & \cdots & 1 \\
        \vdots & \vdots & \ddots & \vdots \\
        a_{r1} & a_{r2} & \cdots & 1
    \end{bmatrix}
    \begin{bmatrix}
       \phi_1 \\ \phi_2 \\ \vdots \\ a_{1r}\phi_r+\cdots+a_{1m}\phi_m
    \end{bmatrix},
\end{equation}
and thus there is no way to determine parameters $\phi_1,\cdots,\phi_{r-1}$.} 

{We note that, when $\bm{v}_{\rm ker}$ has a zero component, there is the case that several local parameters are exposed to untrusted parties. For example, let us consider a sensor network for estimating $f(\bm{\phi})=\phi_1+\phi_2+\phi_3$, whose FIM has one kernel vector $\bm{v}_{\rm ker}=[0 \ \ 1 \ \ -1]$. This enables estimating $g(\bm{\phi})=\phi_1+\frac{1}{2}\phi_2+\frac{1}{2}\phi_3$ within finite variance, allowing $\phi_1$ to be eavesdropped.}

\section*{Appendix C. Discussion for the case of \texorpdfstring{$0<\mathcal{P}_{\mathbf{F}}(\bm{w})<1$}{0<P_F(w)<1}}

{We discuss the physical meaning of intermediate values of the privacy quantifier, $0<\mathcal{P}_{\mathbf{F}}(\bm{w})<1$. Consider a FIM whose support space has dimension $(m-1)$, with the spectral decomposition
\begin{equation}
\mathbf{\Pi}_{\mathbf{F}}=\sum_{j=1}^{m-1}\bm{e}^{(j)}\bm{e}^{(j)\mathrm{T}},
\end{equation}
where $\{\bm{e}^{(j)}\}$ are orthonormal vectors and $\bm{e}^{(m)}$ spans the kernel of $\mathbf{F}$.}

{Let $\bm{v}$ be a unit vector orthogonal to $\bm{w}$, expanded as $\bm{v}=\sum_{j=1}^{m}c_j\bm{e}^{(j)}$. Then
\begin{equation}
    \bm{v}^{\rm T}\mathbf{\Pi}_{\mathbf{F}}\bm{v}=|c_1|^2+|c_2|^2+\cdots+|c_{m-1}|^2,
\end{equation}
If $\bm{w}$ has a nonzero component along the kernel direction $\bm{e}^{(m)}$, the minimizing vector $\bm{v}$ necessarily overlaps with the support space, yielding $\bm{v}^{\mathrm{T}}\mathbf{\Pi}_{\mathbf{F}}\bm{v}<1$ and thus $0<\mathcal{P}_{\mathbf{F}}(\bm{w})<1$.}

{Physically, the kernel direction splits the global function into two components: $(\mathbf{\Pi}_{\mathbf{F}}\bm{w})^{\mathrm{T}}\bm{\phi}$ and $((\bm{e}^{(m)}\bm{e}^{(m)\mathrm{T}})\bm{w})^{\mathrm{T}}\bm{\phi}$. Only the former can be estimated with finite precision~\cite{m.namkung3,p.stoica,y.-h.li}, while the latter remains completely hidden. This regime therefore represents a geometric trade-off: privacy is guaranteed, but only the component of the global function supported by the FIM can be accessed by the clients.}

\section*{Appendix D. Verifying basis independence, continuity, and noise tolerance}

\subsection*{D.1. Verification of basis independence}
{We verify that the universal privacy quantifier $\mathcal{P}_{\mathbf{F}}(\bm{w})$ defined in Eq.~(\ref{pp}) is invariant under changes of basis in the parameter space. Let $\mathbf{O}$ be an orthogonal matrix implementing a basis transformation. Since $\mathbf{O}$ preserves inner products, we have
\begin{align}
    \min_{\substack{\bm{v}\perp\bm{w},\\ \|\bm{v}\|_2=1}}
    \bm{v}^{\rm T}\mathbf{\Pi}_{\mathbf{F}}\bm{v}
    =
    \min_{\substack{\mathbf{O}\bm{v}\perp\mathbf{O}\bm{w},\\ \|\mathbf{O}\bm{v}\|_2=1}}
    (\mathbf{O}\bm{v})^{\rm T}
    (\mathbf{O}\mathbf{\Pi}_{\mathbf{F}}\mathbf{O}^{\rm T})
    (\mathbf{O}\bm{v}).
\end{align}
The projector $\mathbf{\Pi}_{\mathbf{OFO}^{\rm T}}$ associated with the transformed FIM $\mathbf{OFO}^{\rm T}$ is constructed from the rotated eigenvectors of $\mathbf{F}$, implying
$\mathcal{P}_{\mathbf{F}}(\bm{w})=\mathcal{P}_{\mathbf{OFO}^{\rm T}}(\mathbf{O}\bm{w})$.
Therefore, the privacy quantifier is basis independent.}

\subsection*{D.2. Verification of continuity}
{We show that the privacy quantifier is continuous under small perturbations of the FIM.
Let $\{\bm{e}^{(j)}\}$ denote an orthonormal basis of eigenvectors spanning the support space of $\mathbf{F}$, and consider a perturbed FIM $\mathbf{F}(\epsilon)$ whose support is spanned by vectors $\bm{e}^{(j)}+\epsilon\bm{f}^{(j)}$, where $\epsilon\ll1$ and $\bm{f}^{(j)}$ represent small perturbations induced by experimental imperfections. The corresponding projector reads
\begin{eqnarray}
    \mathbf{\Pi}_{\mathbf{F}(\epsilon)}=\sum_j\left(\bm{e}^{(j)}+\epsilon\bm{f}^{(j)}\right)^{\rm T}\left(\bm{e}^{(j)}+\epsilon\bm{f}^{(j)}\right).
\end{eqnarray}
For any unit vector $\bm{v}$, we obtain
\begin{eqnarray}
    \left|\bm{v}^{\rm T}\left\{\mathbf{\Pi}_{\mathbf{F}(\epsilon)}-\mathbf{\Pi}_{\mathbf{F}}\right\}\bm{v}\right|\le \epsilon A+\epsilon^2 B,
\end{eqnarray}
with finite coefficients
$A=2\left|\sum_j(\bm{v}^{\rm T}\bm{e}^{(j)})(\bm{f}^{(j)\rm T}\bm{v})\right|$
and
$B=\left|\sum_j(\bm{v}^{\rm T}\bm{f}^{(j)})^2\right|$.
This establishes the continuity relation in Eq.~(\ref{continuity}).}

\subsection*{D.3. Verification of noise tolerance}
{We show that perfect privacy, $\mathcal{P}_{\mathbf{F}}(\bm{w})=1$, is preserved under noisy channels.
Any parameter-independent noise process transforms the encoded probe state into
$(1-D)\hat{\rho}_{\bm{\phi}}+D\hat{\sigma}$,
where $\hat{\sigma}$ is independent of the parameters and $D\in[0,1]$ quantifies the noise strength.
For a fixed measurement $\mathcal{M}$, the FIM is convex in the state, yielding
\begin{equation}
    \bm{v}^{\rm T}\mathbf{F}\Big((1-D)\hat{\rho}_{\bm\phi}+D\hat{\sigma},\mathcal{M}\Big)\bm{v}\le(1-D)\bm{v}^{\rm T}\mathbf{F}(\hat{\rho}_{\bm\phi},\mathcal{M})\bm{v},
\end{equation}
where $\mathbf{F}(\hat{\sigma},\mathcal{M})=0$ since $\hat{\sigma}$ carries no parameter information.
If $\mathcal{P}_{\mathbf{F}}(\bm{w})=1$, there exists a direction $\bm{v}$ in the kernel of $\mathbf{F}$ such that
$\bm{v}^{\rm T}\mathbf{F}(\hat{\rho}_{\bm{\phi}},\mathcal{M})\bm{v}=0$.
Consequently,
\begin{eqnarray}
    \bm{v}^{\rm T}\mathbf{F}\big((1-D)\hat{\rho}_{\bm\phi}+D\hat{\sigma},\mathcal{M}\big)\bm{v}=0,
\end{eqnarray}
demonstrating that perfect privacy is preserved under parameter-independent noise.}
\end{widetext}

\section*{Appendix E. Experimental details}

 \begin{figure*}[t]

\includegraphics[width=\textwidth]{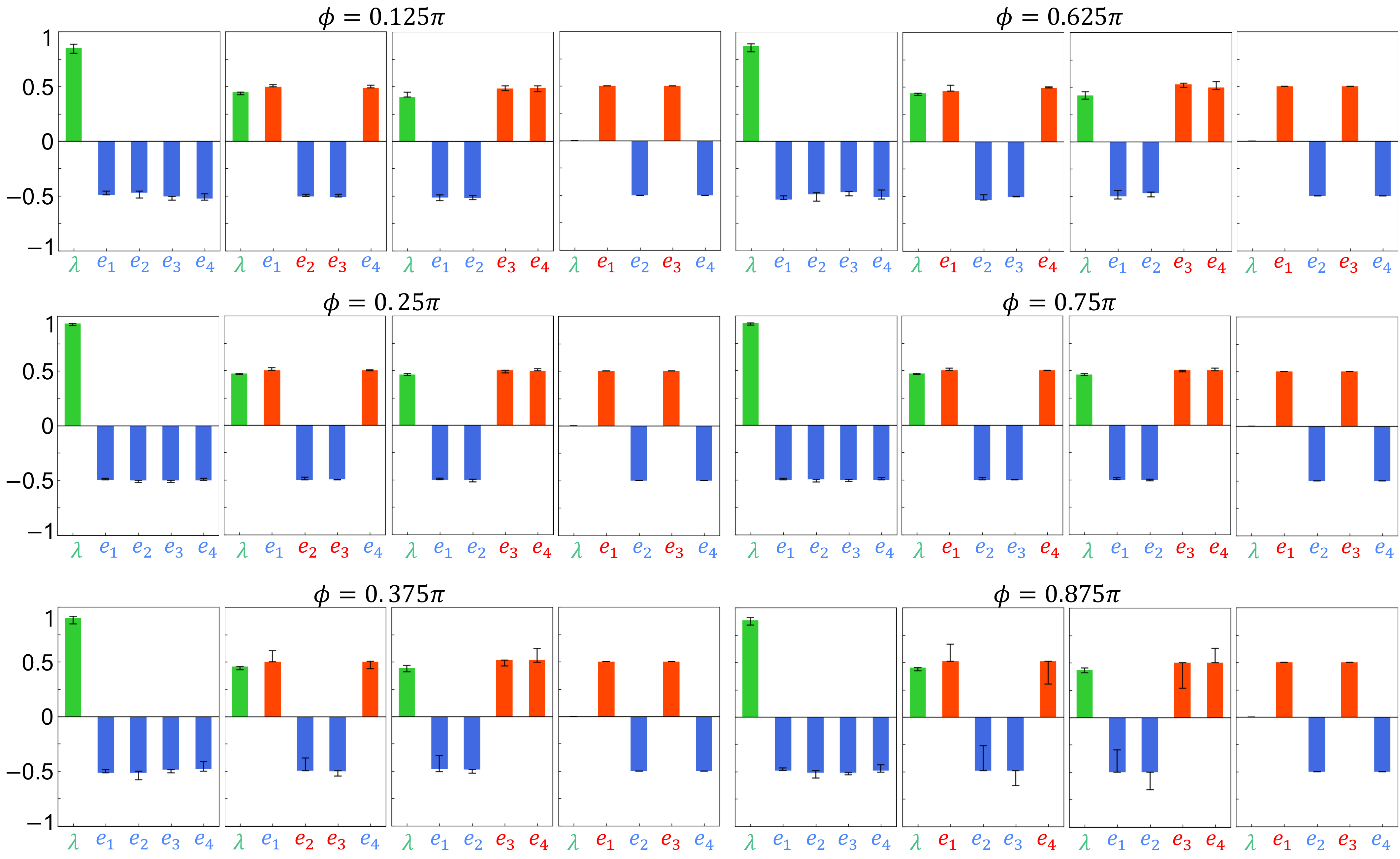} 
\caption{{Four eigenvalues and eigenvectors of the experimentally reconstructed FIMs. The horizontal axis shows the eigenvalues $\lambda$, while the vertical axis displays the corresponding eigenvector components $(e_1,e_2,e_3,e_4)$. In the experiment, all phases are set to a common value $\phi$, i.e., $\phi_1=\phi_2=\phi_3=\phi_4=\phi$. Error bars indicate the 95\% confidence interval.}}
\label{Sfig:1privacy}
\end{figure*}

{Here we provide additional experimental details supporting the demonstration of private distributed quantum sensing for estimating the global function $(\phi_1+\phi_2+\phi_3+\phi_4)/4$. Fig.~\ref{Sfig:1privacy} presents the experimentally reconstructed FIMs, including their four eigenvalues and corresponding eigenvectors.}

{As shown in Fig.~\ref{Sfig:1privacy}, the experimentally obtained FIMs closely reproduce the theoretical FIM, which has eigenvalues $(1,\,0.5,\,0.5,\,0)$ and corresponding eigenvectors $(-0.5,-0.5,-0.5,-0.5)$, $(-0.5,0.5,0.5,-0.5)$, $(0.5,0.5,-0.5,-0.5)$, and $(-0.5,0.5,-0.5,0.5)$~\cite{d.-h.kim}. In particular, the presence of the eigenvector $(-0.5,0.5,-0.5,0.5)$ associated with a zero eigenvalue confirms the singular structure of the FIM realized in the experiment.}

{This experimentally verified kernel structure directly supports the privacy condition discussed in {Appendix B}. Specifically, the existence of a kernel vector orthogonal to balanced global weight vectors implies that no information about any individual local phase is accessible to the untrusted servers. These observations confirm that the distributed quantum sensing network implemented in our experiment simultaneously achieves the desired global estimation task and the operational privacy guaranteed by the universal privacy condition.}


\begin{thebibliography}{99}
\bibitem{footnote}
{Actually, estimation of a single parameter $\phi_j$ is equivalent to that of a function $\bm{e}_j^{\rm T}\bm{\phi}$, where $\bm{e}_j$ denotes the $j$th orthonormal basis vector of the parameter space.}
\end{thebibliography}
\end{document}